# Machine Learning Method to Control and Observe for Treatment and Monitoring of Hepatitis B Virus

SeyedMehdi Abtahi[1], Mojtaba Sharifi[2,1,*]

[1] *Department of Mechanical & Industrial Engineering, University of Illinois at Chicago, Chicago, Illinois, USA*

[2] *Department of Electrical and Computer Engineering, University of Alberta, Edmonton, Alberta, T6G 1H9 Canada*

*\* Corresponding author. Tel.: +1-587-712-4147.  E-mail: sharifi3@ualberta.ca .*

# Machine Learning Method to Control and Observe for Treatment and Monitoring of Hepatitis B Virus

**Abstract**

Hepatitis type B is one of the most common infectious disease worldwide that can pose severe threats to human health up to the point that may contribute to severe liver damage or cancer. Over the past two decades, a large number of dynamic models have been presented based on experimental data to predict the HBV (hepatitis B virus) infection behavior. Besides, several kinds of controllers (linear, optimal, and adaptive) have been employed to obtain effective solutions from the HBV treatment. In this essay, we consider the nonlinear HBV dynamic model, which subjected to both parametric and non-parametric uncertainties, without using any linearization. In previous control methods, three HBV dynamic states should be measured: virus, safe, and infected cells. However, in most of the biological systems, the amount of virus is experimentally measured. Accordingly, the necessity to seek a method that can estimate the amount of required drug by receiving the virus data emerges. An ANFIS (Adaptive Nero-Fuzzy Integrated System) method is developed in this work to provide an intelligent controller for the drug dosage based on the number of viruses together with an estimator (state observer) for the amount of infected and uninfected cells. This controller is trained first using the data provided from a previous adaptive control strategy. After that, to improve the closed-loop system's capabilities, two unmeasured state variables (infected and safe cells) of fundamental dynamics are estimated through the training phase of the ANFIS observer. The results of simulations demonstrated that the accuracy of the proposed intelligent controller is high in the tracking of the desired descending virus population.

## 1. Introduction

According to the latest figures released by WHO (world health organization), more than 2 Billion people have been suffered from Hepatitis. Although only 1% percent of America and Europe were infected in the last ten years, figures suggest that 10% of the population lived in Asia and Africa have been affected (Riberiro RM, Lo A, Perelson AS, 2002), and more than half of these infected people are living in Asia. More importantly, Hepatitis is a leading cause of inevitable deaths (K. Hattaf, M. Rachik, S. saadi, Y. Tabit and N. Yousefi, 2009). Besides, this infection is passed on through the birth process, Sexual contact, and even between children who live in places with poor sanitation (Kapoor. A, Bhatia. V, Gopalan. S and Sibal. A, 2011). Chronic HBV can have adverse effects on human health up to the point that it may cause irreversible damages and even death. As a consequence, governments are expected to take immediate action to address this disease.

On the other hand, available drugs in the market do not enable to solve this problem completely. Nevertheless, they have the ability to decrees and even stop the rate of virus replication. We used the dynamic models, which makes us capable of controlling virus replication and monitoring immune system responses.

The first mathematical modeling of HBV infection's researches was done on animals, including woodchuck HBV (D, 1996) and duck HBV; although there are some fundamental differences between humans and these animals in response to drugs, these researches prepared the ground for finding more precise models. Scientists to gain a profound insight into the nature of this disease should find a model that can describe the dynamics of the virus, safe, and infected cells. Indeed, it suggests that experiments and mathematics models should collaborate and cooperate to find an appropriate mathematical model and serve the interest of patients. These models have used for common diseases, including HVB (Nowak & et al., 1996; Desta & Koya, 2019), HIV (human immunodeficiency virus) (Landi. A, Mazzoldi. A, Andreoni. C, Bianchi. M, Cavallini. A, Laurino. M, Ricotti. L, IUliano. R, Matteoli. B and Ceccherini-Nelli. L, 2008), and HCV (hepatitis C virus) (Debroy & et al., 2010).

The underlying mathematical model used in this paper was introduced by Nowak et al. (Nowak & et al., 1996), which is the most common and accepted model, and a large number of previous researches have used this proposed dynamic. In this study, the first time, 45 patients, and in the second study, 50 patients were treated for 28 days and 24 weeks, respectively. This research was the first step in presenting a mathematical model to deal with HBV. The next significant research was done by Hattaf et al. (K. Hattef, M. Rachik, S.Saadi and N. Yousfi, 2009), which introduced drugs as an input control variable for the first time. Since then, a large number of papers have been published used this model (Riberiro RM, Lo A, Perelson AS, 2002; Sheikhan & Ghoreishi, 2013; Sheikhan & Ghoreishi, 2013). Controlling the number of viruses and reducing the required drugs are the essential concerns of scientists in these researches. However, the other significant information necessitated for specialists contains the amount of safe and infected cells over time. Therefore, requiring a method to monitor these states emerges. These observers can lead to a substantial opportunity for specialists to have comprehensive information about the immune system of each patient during treatment. In the last twenty years, a vast majority of control methods have been used to treat HVB, and a variety of scientists has chosen the optimal way. K. Hatta et al. (K. Hattef, M. Rachik, S.Saadi and N. Yousfi, 2009) studied an optimal controller to demonstrate the efficacy of drug therapy. Jonathan E. Forde et al. (Jonathan E. Ford, Stanca M. Ciupe, Ariel Cintron-Arias and Suzanne Lenhart, 2016) used an optimal controller of the hepatitis B model to reduce the adverse side effects of combination therapy. Su et al. studied the comparison of the impact of two different drugs in the HBV treatment process. In the other work, they developed an optimal method to show the priority of this method over the constant dosage. Laarabi et al. (H. Laarabi, A. Abta, M. Rachik and J. Bouyaghroumni, 2013) used an optimal controller to minimize treatment costs and reduce the volume of healthy cells. Also, various types of controllers have been used to treat other infectious diseases. For example, an optimal controller has been used to control HIV (J.F. De Souza, M.A.L. Ceatano and T. Yoneyama, 2008; U. Ledzewicz and H.Schattler, 2002). Ryan Zurakowski and Andrew R. Teel studied a predictive model control method for HIV therapy (Rayan Zurakowski and Andrew R. Teel, 2006). Alberto Landi et al. (Landi. A, Mazzoldi. A, Andreoni. C, Bianchi. M, Cavallini. A, Laurino. M, Ricotti. L, IUliano. R, Matteoli. B and Ceccherini-Nelli. L, 2008) considered aggression as a new state variable to quantify the strength of the virus and its response to drugs in HIV patients.

In the other example, an optimal controller was used to treat the flu (J. Lee, J. Kim and H.D. Kwon, 2013). Blayne et al. (K. Blayneh, Y. Cao and H.D. Kwon, 2009) studied optimal control of malaria treatment, and their main task was to reduce the number of infected cells. All of these optimal control logics inevitably resulted in a specific mathematical model, leading to mismatch behavior compared to real responses. More importantly, optimal control devices may not be able to predict the exact amount of drug in the presence of uncertainty.

The prevailing view is that the adaptive control method is one of the most impressive models that can address the problem of uncertainty in HVB and other infectious diseases (Aghajanzadeh, et al., 2018). In this context, Sharifi et al. proposed a nonlinear robust adaptive controller in the presence of uncertainty for influenza (Mojtaba Sharifi and Hamed moradi, 2017). In the other study, this team studied a robust adaptive controller for cancer chemotherapy in the presence of parametric uncertainties to manipulate drug usage (Moradi, et al., 2015). Rokhforuze and et al. (Pegah Rokhforoz, Arta A. Jamshidi and Nazanin Namazi Sarvestani, 2017) studied a robust adaptive controller for cancer with different drug users. The other work was done by Aghajanzadeh and et al. (Omid Aghajanzadeh, Mojtaba Sharifi, Shabnam Tashakori and Hassan Zohour, 2017)

Used a nonlinear adaptive control method to treat hepatitis B virus (HBV) infection. We use this model as a fundamental dynamic model for training the ANFIS controller and observer, detailing the entire training process in the following sections. In this proposed model, three state variables were considered, including viruses, safe, and infected cells. Besides, several methods were used in this study to achieve the correct behavior of the controller in the presence of uncertainties. All of these adaptive control logics are model-based, and although they can work with uncertainties, they do not apply to different models or patients.

No one can dispute the enormous ability of fuzzy logic (Zadeh, 1965) to deal with various problems. Also, fuzzy logic is not a model base that helps solve many issues, although the dynamic model may not be unique. Mansour Sheikhan and S. Amir Ghoreishi (M. Shekhan and S.A.Ghoreishi, 2013) used a fuzzy controller for the treatment of HVB. Aboul Ella Hassanien (Hassanein, 2007) applied fuzzy logic to detect breast cancer. Rafayah Mousa et al. (Rafayah Mousa, Qutaishat Munib and Abdallah Moussa, 2005) employed a fuzzy-neural method to diagnose breast cancer. In other researches, Carlos Andrés Peña-Reyes and Moshe Sipper used fuzzy-genetic for diagnosing breast cancer (Carlosn Andres, Pena-Reyes, Moshe Sipper, 1999). Hassan Zarei et al. Fuzzy logic has been used to treat HIV infection by considering various mathematical models over the past decade (Hassan Zarei, Ali Vahidan Kamyad and Ali Akbar Heydari, 2010; Hassan Zarei, Ali Vahidian Kamyad, Sohrab Effati, 2011; Jafelice, et al., 2019).

In this paper, an ANFIS (adapted neuro-fuzzy inference system) (Jang, 1993), the controller is introduced to treat the infection hepatitis B virus (HBV), which can work in the presence of unstructured uncertainty. The most important feature of this ANFIS controller is the ability to handle the free model system, which allows designing a controller based on the provided real data. This superior ability is naturally followed by eliminating the erroneous may exist in mathematical models. The other considerable superiority of the ANFIS system is decreasing the necessitate number of inputs variable to estimate the appropriate amount of drug. Although previously suggested controllers for the HBV at least required three Input variables including viruses, safe and infected cells, ANFIS controllers indicated in this paper need a virus as an input variable. Besides, the ability to observe the other necessary states of previous control strategies, safe and infected cells, introduces this model as a significant model, given the inability to use an observer and an adaptive controller at the same time. Also, the previously proposed controller required feedback from all states, and here the importance of a practical observer comes into play. Therefore, for the first time, the proposed ANFIS system provides all the crucial data necessary to treat HBV infection.

This paper is organized as follows. In section 2, the underlying mathematical model used is introduced. ANFIS structure is discussed entirely in Sec. 3. Then, it outlines how the ANFIS controller is designed in Sec. 4. The previous section repeated this time for ANFIS observer in Sec. 5. The simulation's results are demonstrated and compared, and the conclusion is expressed in Sec. 6

## 2. Mathematical Model of HBV

The basic mathematical model employed in this research, originally proposed by Nowak et al (M.A.Nowak, S. Bonhoeffer, A.M.

Hill, R. Boehme, H.C. Thomas and H. McDade, 1996)(which a vast majority of patients during their treatment interval were scrutinize), considers safe cells (x), infected cells (y) and virus (v) using the following equations:

$$\frac{dx}{dt} = l - dx - bvx \quad\quad\quad 1$$

$$\frac{dy}{dt} = (1-u)bvx - dy \quad\quad\quad 2$$

$$\frac{dv}{dt} = py - cv \quad\quad\quad 3$$

"u" is control input express the rate of drug usage added to model by Hattaf et al. (K. Hattef, M. Rachik, S.Saadi and N. Yousfi, 2009). "b, d, c, p and $l$ are model parameters which were represented again by Hattaf et al (K. Hattef, M. Rachik, S.Saadi and N. Yousfi, 2009) by all details which are represented in table 1.

**Table 1** Parameters value

| Parameters | Value |
|---|---|
| $l$ | 1/3 |
| d | 6.28947 |
| b | 1/3 |
| p | 767.610/3 |
| c | 376.3157/3 |

## 3. Adaptive network-based fuzzy inference system
### 3.1 Fuzzy
Traditional methods of modeling and arithmetic are precise. It means that a statement can be true or false and nothing in between. L.A.Zadeh (Zadeh, 1965) proposed a fuzzy logic that is "partially true" for the first time. Blurs can be observed in a variety of areas of daily life, especially in engineering (Blockley, 1980) and medicine (M.A. Vila and M. Delgado, 1983), and so on. In fuzzy logic input and output, variables can take on continuous values between 0 and 1. First time Mamdani and Assilian (E.H. Mamdani and S. Assilian, 1975) used fuzzy logic in technical problems. A fuzzy controller designer writes rules to link the input variables to the outputs via linguistic variables.
After defining the rules, the calculation of all control sequences can be started and a fuzzy set describing the possible control actions.
The fuzzy control logic has two main approaches:
Mamdani (E.H. Mamdani and S. Assilian, 1975): figure1 depicts a Mamdani fuzzy controller, and each part will be discussed later.

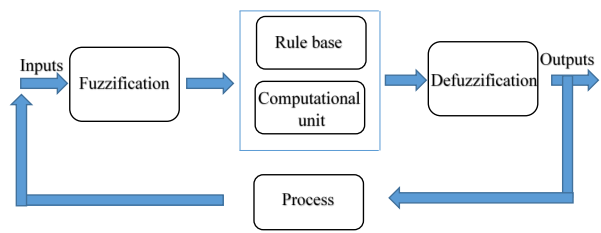

**Fig 1.** Schematic diagram of Mamdani fuzzy control logic.

Input: number of input signals, number of derived states of each input signal, scaling of the input signal.

Fuzzification: type of membership function, mean, spread, and the peak of membership functions, symmetry, cross points, continuous or discrete support.

Rules: number of rules, number of antecedents, the structure of rule base, type of membership functions in consequences, rule weights.

Rule evaluation: aggregation operator in the antecedent, inference operator.

Aggregation: aggregation operator combining the results of the individual rules, own rule-based inference, or composition-based inference.

Defuzzification: defuzzification process in the opposite of fuzzification.

Output: number of output signals

Takagi-Sugeno *(Michio SUGENO and Tomohiro TAKAGI, 1985)*: the Takagi-Sugeno fuzzy model, initially proposed by Takagi and Sugeno. The consequent of each rule is a purely functional expression; this approach is the same as the Mamdani method with a difference in the output type. In this model, the output is a number or a linear function of input variables.

*If* $z_i$ is $Z_1^i$ and ... and $z_p$ is $Z_p^i$ Then y= $F_i(z)$
Where the vector z has p components.

### 3.2 ANFIS

Takagi-Sugeno fuzzy system (Michio SUGENO and Tomohiro TAKAGI, 1985) is the base of the ANFIS (Adaptive Network-Based Fuzzy Integrated system) (Jang, 1993) method. Its inference system corresponds to a set of fuzzy IF-THEN rules that have a learning ability to approximate non-linear functions. ANFIS first defines the membership function of the inputs and then estimates the possibility of each rule. Later try to normalize the output from the previous step. The normalization and the linear function of the input variables can cause these results. The last problem is the sum of all of the issues is layer 4.

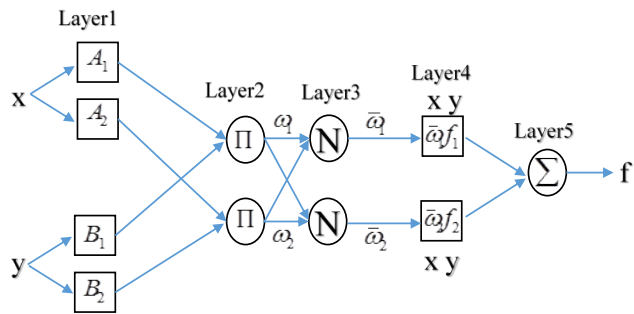

**Figure 2** ANFIS Structure

ANFIS has 5 layers (figure 1) as follows:
Define membership function

$$O_{1,i} = m_{A_i}(x) \quad for\ i = 1,2$$
$$O_{1,i} = m_{B_i}(x) \quad for\ i = 3,4$$

Product of the membership function for each input.

$$O_{2,i} = \omega_i = m_{A_i}(x)m_{B_i} \quad i = 1,2$$

Normalize the output of layer 2.

$$O_{3,i} = \bar{\omega}_i = \frac{\omega_i}{\omega_1 + \omega_2} \quad i = 1,2$$

Then in layer 4:

$$O_{4,i} = \bar{\omega}_i f_i = \bar{\omega}_i(p_i x + q_i y + r_i)$$

summation of all outputs in layer 4 would be final answers.

$$O_{5,i} = \sum \overline{\omega}_i f_i = \frac{\sum \omega_i f_i}{\sum \omega_i}$$

### 3.2.2 ANFIS controller

The purposed ANFIS controller allows estimating the drug by looking at the virus as an input variable. To this end, learning data is provided that exploits various uncertainties of the dynamic base model and each time stores viruses and drugs as input and output variables. The Gaussian MF type is selected, and it is preferred to output it as a linear function of the input to be more accurate. Finally, the ANFIS controller training process begins, and the ANFIS system learns to evaluate outputs.

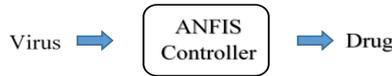

**Figure 3** ANFIS Control Diagram

Figure 3 shows that the optimal amount of drug during the prediction will be predicted. We just need to have some information about the Virus and then our proposed model can predict the optimal amount of drug during the treatment.

### 3.2.3 ANFIS observer

In this paper, an ANFIS observer is connected to the controller to estimate the other necessary states of the fundamental dynamics system. Since the behavior of safe cells, as we see in equation (1), depends on the virus to train ANFIS observers to estimate safe cells, the virus is considered to be an input variable, and the output variable is safe cells.

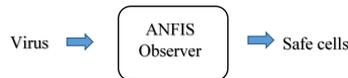

**Figure 4** ANFIS Observer to estimate safe cells

To estimate the number of infected cells by ANFIS observers with reference to equation (2), it is obvious that viruses and safe cells should be input variables and infected cells should be considered as starting variables.

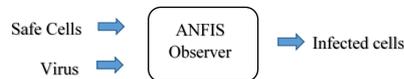

**Figure 5** ANFIS Observer to estimate infected cells

To estimate the number of infected cells by ANFIS observers with reference to equation (2), it is obvious that viruses and safe cells should be input variables and infected cells should be considered as starting variables.

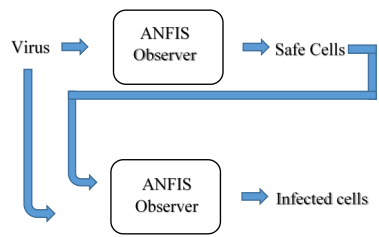

**Figure 6** ANFIS Observer total structure

### 4. Results

In this section, two different scenarios (one patient with two various uncertainties) were considered and simulated to confirm the capabilities of the ANFIS controller and observer. Also, we compare the results of the ANFIS controller with the adaptive base controller and the effects of ANFIS observers with the actual data available.

To study the performance of ANFIS controllers and observers, several simulations were performed. Initial conditions are shown in Table2.

**Table 2** initial values

| parameters | Value |
|---|---|
| X(0) | 0.836057 |
| Y(0) | 0.1672 |
| V(0) | 0.3289294 |

### 4.1 ANFIS Controller result

To demonstrate the ability of the ANFIS controller designed and engineered to track hepatitis B virus infectious, simulation results of ANFIS and ADAPTIVE controllers in one figure were compared. The first graph compares the amount of drug received from the ANFIS and ADAPTIVE controls. In this case, we considered an uncertainty of 20%. Besides, this condition was one of the requirements for training ANFIS controls.

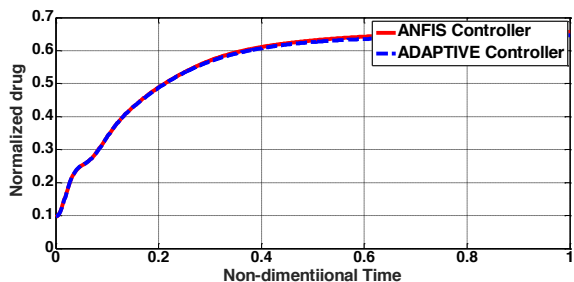

**Figure 7** Amount of drug for patient with 20 percent of uncertainty

It is obvious that the behavior of the ANFIS controller is very similar to the ADAPTIVE controller and the error of these two values is less than 2%.

The next graph compares the amount of medication in these two controllers, but this time this condition was not one of the ANFIS training conditions, along with the randomly selected uncertainty.

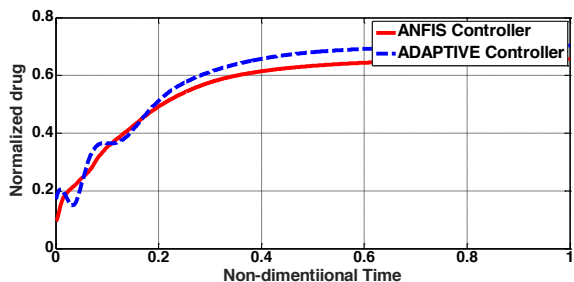

**Figure 8** Amount of drug for patient with random percent of uncertainty

In this case, the ANFIS control exceeds the ADAPTIVE control because the output of the ANFIS controller is smooth and does not fluctuate, and eventually the amount of drug used in this procedure is lower.
In the next image, the area where the output of the ADAPTIVE knob fluctuated was zoomed to best compare the behavior of two controllers.

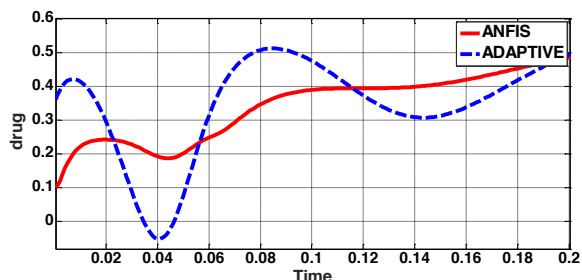

**Figure 9** Amount of drug for patient with random percent of uncertainty (zoom the fluctuate part)

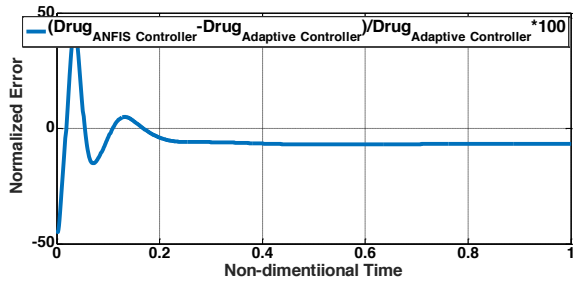
**Figure 10** error of ANFIS in comparison with Adaptive controller drug for patient with random percent of uncertainty

To demonstrate the high capability and ability of ANFIS controllers over ADAPTIVE controllers, the ANFIS controller only needs to use the virus as an input variable, while the ADAPTIVE controller needs viruses, safe and infected cells as input variables. Besides, the ANFIS controller is not model-based and can work with real data, along with a reduction in drug delivery during treatment by more than 6 percent, which is understandable in Figure 8.

### 4.2 ANFIS observer results
#### 4.2.1 Monitoring infected cells
The next two figures show and compare the trend of the dimensionless number of infected cells during treatment using FUZZY observer and input data.

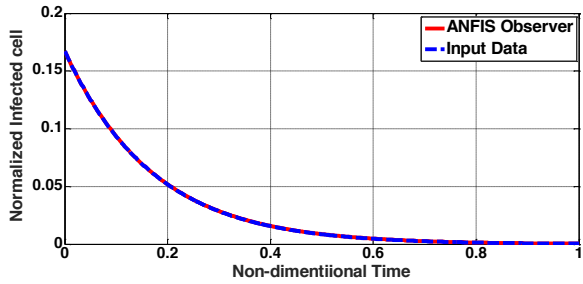
**Figure 11** Number of infected cells for patient with 20 percent of uncertainty

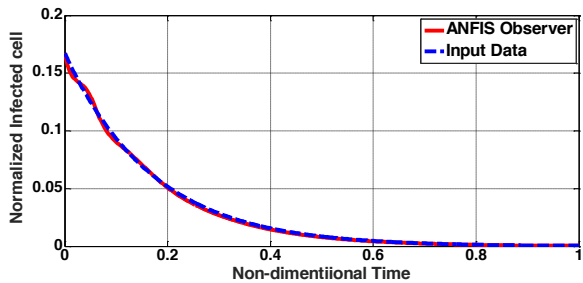
**Figure 12** Number of infected cells for patient with random percent of uncertainty

Results exhibit the high ability of ANFIS observer to estimate the trend of infection cells during treatment precisely in the course of time.

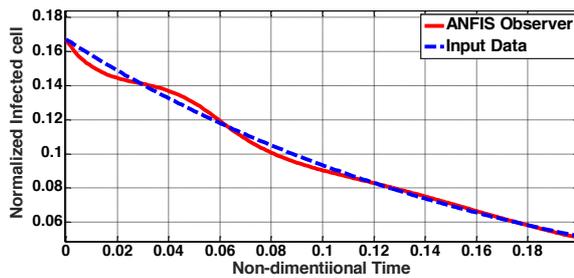
**Figure 13** Number of Infected Cells for patient with random percent of uncertainty (zoom the fluctuate part)

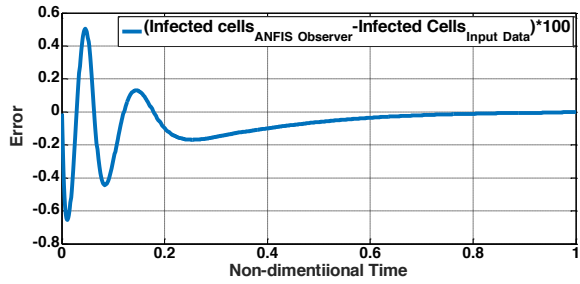
**Figure 14** error of ANFIS Observer in comparison with Input Data of Infected Cells for patient with random percent of uncertainty

These figures show the ANFIS observer's ability to estimate very accurately after a short time, and then it is obvious that the error in the estimate converges to zero.

**4.2.2 Monitoring safe cells**
The next two figures show and compare the trend of the dimensionless numbers of safe cells during the treatment using FUZZY observer and input data, which is an essential help for specialists, while it is impossible to use observers with an ADAPTIVE controller.

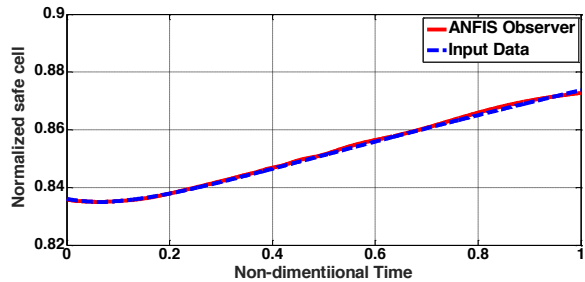
**Figure 15** Number of safe cells for patient with random percent of uncertainty

The above diagrams demonstrate the high ability of ANFIS observer to monitor the other necessary state of controlling system which is really useful and applicable for specialists as they can observe and anticipate the trend of safe and infected cells after each treatments session.

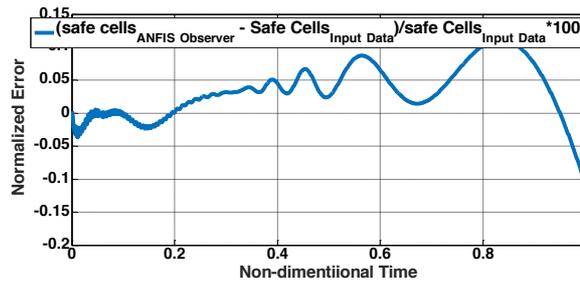
**Figure 16** error of ANFIS Observer in comparison with Input Data of Safe Cells for patient with random percent of uncertainty

This figure shows that although the error fluctuate and does not converging, the percentage of error is very infinitesimal and negligible. The Maximum amount of error is about 0.1 percent.

# 5. Conclusion
In this study, an ANFIS (adaptive neuro-fuzzy inference system) controller was presented for treatment of the hepatitis B virus (HBV) infection along with rendering an ANFIS observer to estimate necessary states of the system. ANFIS controller only needs the amount of illness as an input variable to determine the amount of requiring drugs, which leads to establishing a connection between medical and engineer science. Therefore, we used real data obtained from patients to train the ANFIS controller directly. The other significant ability of the system presented in this paper was estimating safe and infected cells, where the only given variable was a virus. However, in previous models, at least three necessary states should be known to estimate the amount of drug, and more importantly, using observer was impossible.

In simulations, one patient with two different conditions (uncertainty) was considered. We compared the performance of the ANFIS controller with the adaptive controller in the presence of various uncertainties. On the other part, the performance of compare the result of ANFIS observer with the real data and show the accuracy and the high ability of it.

Hopefully, ANFIS controller and observer will be able to work with discrete data (amount of drug) to be employed in realistic clinical treatments, which contribute to estimating the amount of medicine that should be injected per day or week. Moreover, specialists can predict the amount of safe and infected cells after each time injecting the drug.